\documentclass[twocolumn]{webofc}

\usepackage[varg]{txfonts}   
\usepackage{hyperref}
\usepackage{url}
\hypersetup{colorlinks=true,citecolor=blue,urlcolor=blue,linkcolor=blue}
\usepackage{physics}
\usepackage{xcolor}

\begin{document}
\title{Constraining of Nuclear Matter Equations of State \\ With Rotating Neutron Stars}
%
%

\author{\firstname{Hyukjin} \lastname{Kwon}\inst{1}\fnsep\thanks{\email{kwon.h.04c4@m.isct.ac.jp}} \and
        \firstname{Kazuyuki} \lastname{Sekizawa}\inst{1,2,3}\fnsep\thanks{\email{sekizawa@phys.sci.isct.ac.jp}} 
}

\institute{Department of Physics, School of Science, Institute of Science Tokyo, Tokyo 152-8550, Japan
\and
           Nuclear Physics Division, Center for Computational Sciences, University of Tsukuba, Ibaraki 305-8557, Japan
\and
           RIKEN Nishina Center, Saitama 351-0198, Japan
          }

\abstract{Neutron stars can be regarded as natural laboratories that enable us to investigate nuclear matter properties under extreme conditions that are otherwise impossible to access in terrestrial experiments. Astrophysical observations of neutron stars provide invaluable information on existing nuclear interaction models and equations of state (EoSs) at various densities. Most studies of neutron star structure employ the Tolman-Oppenheimer-Volkoff (TOV) equation which describes spherically symmetric, non-rotating stars in hydrostatic equilibrium. However, since neutron stars rotate fast, they could experience significant centrifugal deformation, and axially-symmetric calculations are required for accurate description of internal structure. The Komatsu-Eriguchi-Hachisu (KEH) method is well known for modeling rapidly-rotating compact objects in a fully general relativistic manner. In this contribution, we report results of KEH calculations for rapidly-rotating neutron stars using EoSs based on Gogny-type finite-range effective nucleon-nucleon interactions. Our results show that the mass-radius relation systematically changes with increasing angular velocity, highlighting the importance of including rotational effects when confronting theoretical EoSs with observational data.
}
\maketitle

\section{Introduction}

A neutron star serves as an excellent testing ground for studying nuclear interactions at various densities. By constructing an equation of state (EoS) based on nuclear interaction models, we can calculate the neutron star structure. This process allows us to refine these models, through comparisons with existing observational data, commonly referred to as constraining the nuclear matter EoS. The most frequently referenced observational data in such constraining studies come from pulsars of PSR J0030+0451 (205\,Hz)~\cite{Miller2019}, PSR J0740+6620 (346\,Hz)~\cite{Wolff2021}, and PSR J1748-2446ad (716\,Hz)~\cite{Hessels2006}, which are rotating rapidly. The observed masses and radii of these neutron stars allow us to make direct comparisons with theoretical predictions.

However, most studies of neutron stars in nuclear physics community are based on the Tolman-Oppenheimer-Volkoff (TOV) equation~\cite{Oppenheimer1939}, which assumes a spherically symmetric, hydrostatic equilibrium (\textit{i.e.} non-rotating) configuration. The latter approach cannot account for deformation of rotating neutron stars induced by centrifugal forces.
To model rapidly-rotating neutron stars in Newtonian gravity, the Hachisu Self-Consistent Field (HSCF) method~\cite{Hachisu1986} was proposed, which employs the integral form of the Poisson equation which is solved in an iterative manner. Soon after that, the HSCF method was extended to deal with general relativity, commonly referred to as the Komatsu-Eriguchi-Hachisu (KEH) method~\cite{Komatsu1989a}.

Recently, we have newly developed our own computational code for describing structure of rapidly-rotating neutron stars based on the KEH method~\cite{Kwon2025a}. In this contribution, we report unpublished results of KEH calculations with EoSs based on Gogny-type finite-range effective nucleon-nucleon interactions~\cite{Gogny1980}. We discuss properties and some caveats of Gogny EoSs and show corresponding mass-radius relations of rotating neutron stars. Finally, a range of rotational frequencies for constraining nuclear matter EoSs is discussed.
\vspace{-0.2cm}
\section{Equation of State (EoS)}
\label{sec-1}

\subsection{Effective nucleon-nucleon interactions}
\label{sec-2}

In this work, We employ Gogny EoSs which are based on Gogny-type finite-range effective nucleon-nucleon interactions. The form of Gogny interactions reads
\begin{equation}
    \begin{split}
    v(\boldsymbol{r}_1,\boldsymbol{r}_2) &= \sum_{i=1}^{2}(W_i+B_iP_{\sigma}-H_iP_{\tau}-M_iP_{\sigma}P_{\tau})\,e^{-\frac{(\boldsymbol{r}_1 - \boldsymbol{r}_2)^2}{\mu_{i}^{2}}} \\   &\quad + t_0 (1 + x_0 {P}_\sigma)\delta(\boldsymbol{r}_1 - \boldsymbol{r}_2)\rho^{\alpha}\left({\boldsymbol{r}_1+\boldsymbol{r}_2\over{2}}\right)  \\
    &\quad + iW_0({\boldsymbol{\sigma}}_1+{\boldsymbol{\sigma}}_2) \cdot \bigl\{ \boldsymbol{k}' \times \delta(\boldsymbol{r}_1 - \boldsymbol{r}_2) {\boldsymbol{k}} \bigr\},
    \label{eq:Gogny}
    \end{split}
\end{equation}
where $P_\sigma$ and $P_\tau$ denote the spin and isospin exchange operators, respectively. The nuclear matter EoS can be obtained from the Gogny interaction. Details of the derivation and its numerical implementation can be found in Ref.~\cite{Gonzalez2017}.
\vspace{-0.2cm}

\subsection{Nuclear Matter EoS}

In this work, we adopt seven Gogny parameter sets: D1, D1S, and D280 \cite{BLAIZOT1995}; D1M \cite{Goriely2009}; D1M* \cite{GONZALEZBOQUERA2018}; D1N \cite{CHAPPERT2008}; and D1P \cite{Farine1999}. For the neutron star crust, we use the BPS-BBP EoS~\cite{Baym1971}, which is one of the most widely used EoS for the crust region. In the core, the EoS is determined under the condition of $\beta$-equilibrium for $npe\mu$ matter.

\begin{figure}[t]
    \centering
    \setlength{\abovecaptionskip}{-10pt} 
    \includegraphics[width=1\linewidth]{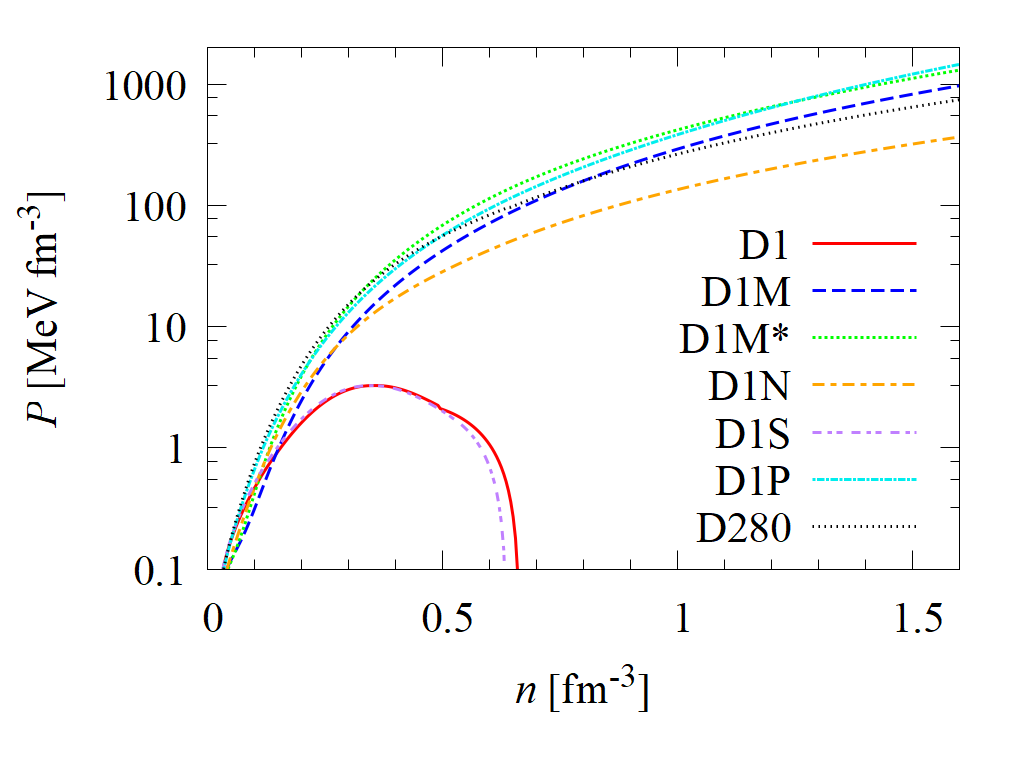}
    \caption{Pressure as a function of baryon (nucleon) number density for $npe\mu$ matter calculated with seven Gogny-type effective nucleon-nucleon interactions examined.}\vspace{-3mm}
    \label{fig:eos}
\end{figure}

In Fig.~\ref{fig:eos}, we show EoSs for $npe\mu$ matter with the seven Gogny EoSs. We find that the six EoSs except D1M$^*$ provides \textit{negative} symmetry energy at certain high density region, meaning that the matter actually becomes the pure neutron matter (PNM). Therefore, in the following discussions, we show results obtained with D1M*.

\section{Models of Neutron Star Structure}

The structure of a hydrostatic object is determined by the balance between pressure and gravity. The equilibrium condition can be expressed as a Poisson equation:
\begin{equation}
    \nabla P = -\rho \nabla \Phi,
    \label{eq:poisson}
\end{equation}
which is a one-dimensional ordinary differential equation that can easily be solved with various computational methods.

However, when the object is rotating, the centrifugal force and deformation must be considered. Equation~\eqref{eq:poisson} then becomes
\begin{equation}
    \nabla P = -\rho \nabla \Phi + \rho \Omega^2 \varpi\hat{\varpi} ,
    \label{eq:rotatingpoisson}
\end{equation}
where $\Omega$ is the angular velocity and $\varpi$ denotes the cylindrical radius. Since Eq.~\eqref{eq:rotatingpoisson} depends on both the radius and polar angle, it is not trivial how to solve the equation numerically. In the HSCF method, Eq.~\eqref{eq:rotatingpoisson} is reformulated into the following integral form:
\begin{equation} 
\int \frac{1}{\rho} \dd P + \Phi - \int \Omega^2 \varpi d\varpi = C,
\label{eq:intpoi}
\end{equation}
together with the integral form of the Poisson equation,
\begin{equation} 
    \Phi = - G \int \frac{\rho(\boldsymbol{r}')}{|\boldsymbol{r}-\boldsymbol{r}'|}\dd\boldsymbol{r}'. 
\end{equation}

Equation~\eqref{eq:intpoi} can be calculated using self-consistent field iterations. For compact massive objects, general relativity must be considered. The Einstein field equation is given by
\begin{equation} G_{\mu\nu} = R_{\mu\nu} - \frac{1}{2} g_{\mu\nu} R = 8\pi G T_{\mu\nu}. \label{eq:Einstein}
\end{equation}
with the energy–momentum tensor for a perfect fluid,
\begin{equation}
    T^{\mu \nu} = (\varepsilon + P)u^{\mu}u^{\nu} + Pg^{\mu \nu},
    \label{EMtensor}
\end{equation}
where $\varepsilon$ is the energy density, $P$ is the pressure, and $u^\mu$ is the four-velocity.

\subsection{Structure of Static Neutron Stars}

The metric for a spherically symmetry is given by
\begin{equation}
    \dd s^2=-e^{2\nu(r)}\dd t^2+e^{2\lambda(r)}\dd r^2+r^2(\dd\theta^2+\sin^2{\theta}\dd\phi^2),
    \label{eq:TOVmetric}
\end{equation}
where $\nu(r)$ and $\lambda(r)$ are metric potentials. By solving Eqs.~\eqref{eq:Einstein} and \eqref{EMtensor} with the metric of Eq.~\eqref{eq:TOVmetric}, the following set of equations can be obtained:
\begin{equation}
    \begin{aligned}
        &\frac{1}{r^2}e^{-2\lambda}(2r\partial_r\lambda-1+e^{2\lambda})=8\pi G \rho, \\
        & \frac{1}{r^2}e^{-2\lambda}(2r\partial_r\lambda+1-e^{2\lambda})=8\pi G \rho, \\
        & e^{-2\lambda}\bigg[ \partial^2_r\nu  +(\partial_r\nu)^2-\partial_r\nu\partial_r\lambda+\frac{1}{r}(\partial_r\nu-\partial_r\lambda)\bigg] = 8\pi G\rho.
    \end{aligned}
\end{equation}
From the conservation of energy momentum, $\nabla_\mu T^{\mu \nu}=0$, these equations can be reduced to the following coupled differential equations:
\begin{subequations}
    \begin{align}
        \frac{\dd P}{\dd r} &= - (\varepsilon+P)\frac{M+4\pi r^3P}{r(r-2M)},\\
        \frac{\dd M}{\dd r} &= 4\pi r^2 \varepsilon.
    \end{align}
\end{subequations}
which are known as the TOV equations. 

\begin{figure*}[t!]
\centering
\vspace*{1cm}       
\includegraphics[width=\textwidth,clip]{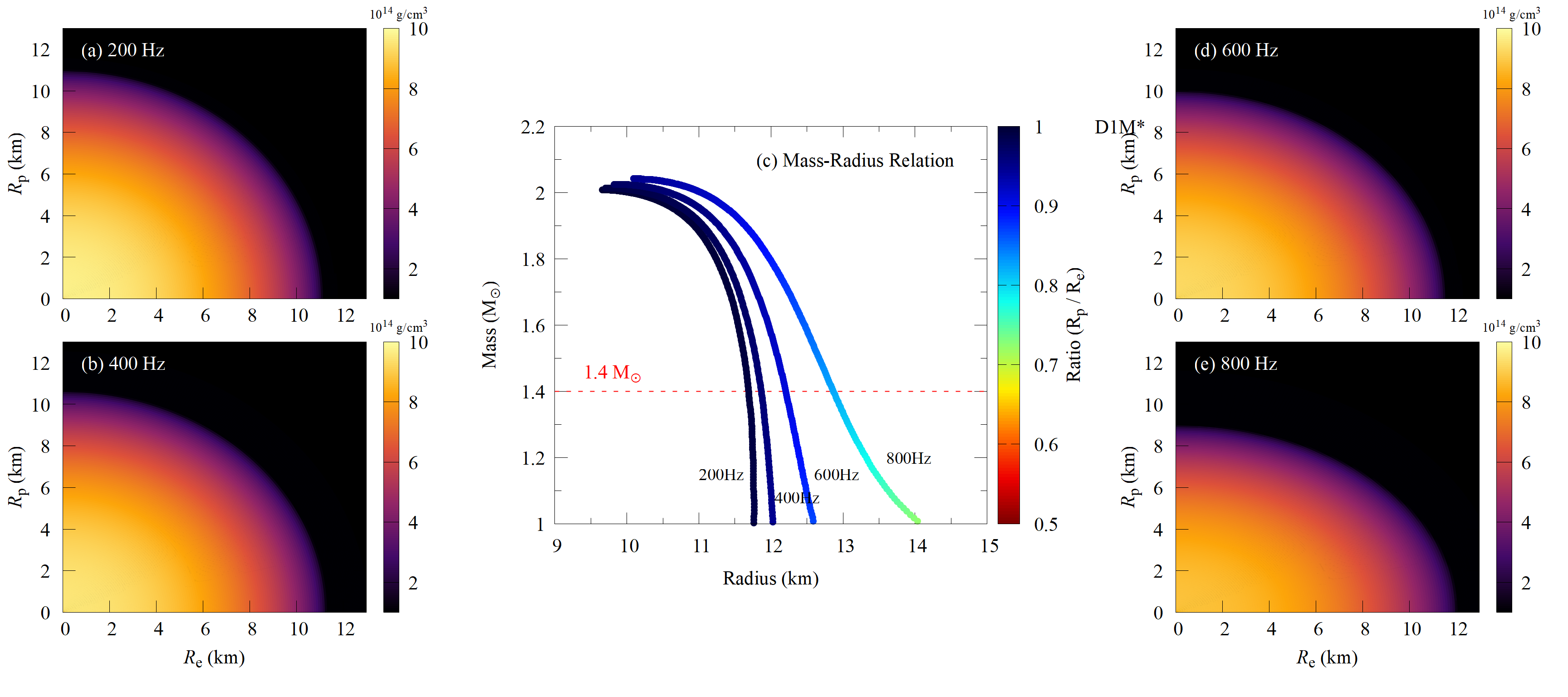}\vspace{-5mm}
\caption{(a--b, d--e) Density distributions of $1.4 \ M_\odot$ neutron stars calculated with the Gogny D1M* equation of state (EoS) at rotational frequencies of 200, 400, 600, and 800\,Hz.
The color scale represents the mass density in units of 
$10^{14} \ \text{g/cm}^3$.
As the rotational frequency increases, the stellar shape becomes more and more oblate due to centrifugal deformation.
(c) Mass-radius relations of neutron stars obtained using the Gogny D1M* EoS at different rotation frequencies.
The color of each curve denotes the ratio of the polar to equatorial radii, $R_{\text{ratio}}=R_\text{p}/R_\text{e}$.
Faster rotation leads to an increase in both the maximum mass and the equatorial radius.}\vspace{-3mm}
\label{fig-2}       
\end{figure*}

\subsection{Structure of Rotating Neutron Stars}

The metric for a axisymmetry spacetime is given by
\begin{equation}
    \begin{aligned}
        \dd s^2 &= -e^{\gamma+\varrho}\dd t^2 + e^{2\alpha}(\dd r^2 + r^2\dd\theta^2) \\
        & \quad + e^{\gamma-\varrho}r^2 \sin^2\theta(\dd\phi-\omega \dd t)^2,
    \end{aligned}
    \label{eq:KEHmetric}
\end{equation}
where $\gamma,\varrho,\alpha, \text{and} \ \omega$ are metric potentials. In the KEH scheme, by solving Eqs.~\eqref{eq:Einstein} and \eqref{EMtensor} together with the metric of Eq.~\eqref{eq:KEHmetric}, one obtains the following Einstein equations:
\begin{equation}
    \begin{aligned}
        &\nabla^2[\varrho e^{\gamma/2}] = S_{\varrho}(r,\mu), \\
        &\left( \nabla^2 + \frac{1}{r} \partial_r - \frac{\mu}{r^2} \partial_{\mu} \right) [\gamma e^{\gamma/2}] = S_{\gamma}(r,\mu), \\
        &\left( \nabla^2 + \frac{2}{r} \partial_r - \frac{2\mu}{r^2} \partial_{\mu} \right) [\omega e^{(\gamma-2\varrho)/2}] = S_{\omega}(r,\mu).
    \end{aligned}
    \label{KEHEQ}
\end{equation}

Equation~\eqref{KEHEQ} can be transformed into integral form using Green's functions:
\begin{equation}
    \begin{aligned}
        &\varrho = -\frac{1}{4\pi}e^{-\frac{\gamma}{2}}\int^\infty_0\dd r'\int^1_{-1}\dd\mu'\int^{2\pi}_0\dd\phi'r'^2S_{\varrho}(r',\mu')\frac{1}{|r-r'|}, \\
        &r\sin\theta\gamma =\frac{1}{2\pi}e^{-\frac{\gamma}{2}}\int^\infty_0\dd r'\int^{2\pi}_0\dd\theta'r'^2 \sin\theta' S_{\gamma}(r',\theta')\log{|r-r'|}, \\
        &r\sin\theta\cos\phi \ \omega = -\frac{1}{4\pi}e^{-\frac{2\varrho-\gamma}{2}}\int^\infty_0\dd r'\int^\pi_{0}\dd\theta'\int^{2\pi}_0\dd\phi' \\
        &\qquad \qquad \qquad \quad r'^3 \sin^2\theta' \cos\phi' S_{\omega}(r',\theta')\frac{1}{|r-r'|}.
    \end{aligned}
    \label{KEHintegral}
\end{equation}
From the conservation law $(\nabla_\mu T^{\mu \nu}=0)$, the condition for equilibrium is given by
\begin{equation}
    \ln{H} + \frac{\varrho+\gamma}{2} + \frac{1}{2}\ln{(1-v^2)} + \int j(\Omega)\dd\Omega = C,
    \label{eqcon}
\end{equation}
where $H$ is the enthalpy and $v$ is the local rotational velocity. This equation is solved iteratively in a self-consistent manner to obtain equilibrium configurations of rotating neutron stars.

\section{Results}

Figure~\ref{fig-2} (c) shows the mass-radius relations of neutron stars rotating at 200\,Hz, 400\,Hz, 600\,Hz, and 800\,Hz, obtained from KEH calculations with Gogny D1M* EoS. As the rotation frequency increases, both the maximum mass and equatorial radius of the neutron star increase.
Furthermore, the ratio between the polar and equatorial radii, $R_{\text{ratio}}=R_\text{p}/R_\text{e}$, decreases with faster rotation, indicating stronger oblateness.
For instance, $R_{\text{ratio}}=0.99$ at 200\,Hz, whereas it decreases to $R_{\text{ratio}}=0.82$ at 800\,Hz.
Even among stars rotating at the same frequency, less massive neutron stars exhibit greater deformation.

For the canonical neutron star mass of $1.4 \ M_\odot$, the density distributions for rotational frequencies of 200, 400, 600, and 800\,Hz are plotted in Fig.~\ref{fig-2}(a), (b), (d), and (e), respectively. In the 200-Hz case [Fig.~\ref{fig-2}(a)], the neutron star maintains an almost spherical shape. As the rotational frequency increases [Figs.~\ref{fig-2}(b), (d), (e)], the polar radius becomes smaller, and the equatorial radius becomes larger, leading to pronounced oblate deformation of the neutron star. A significant deformation can be found at the 800-Hz case shown in Fig.~\ref{fig-2}(e).

The corresponding central densities decrease with increasing rotation frequency.
For instance, in the case of a $1.4\,M_\odot$ neutron star, the central density is 
$9.7\times 10^{14}\ \text{g/cm}^3$ at 200\,Hz, while it decreases to 
$8.8\times 10^{14}\ \text{g/cm}^3$ at 800\,Hz. 
For comparison, the non-rotating TOV solution gives 
$9.8\times 10^{14}\ \text{g/cm}^3$, which is only slightly higher than the value at 200\,Hz. 
This indicates that the internal structure of the star at 200\,Hz remains very close to the spherical, non-rotating configuration.  In contrast, the 800-Hz case exhibits both a substantially reduced central density and a markedly-deformed oblate shape, demonstrating that rapid rotation leads to significant modifications of the neutron star structure. We mention here that even in the case of 200\,Hz, a non-zero deformation of about 1\% relative to a perfect sphere is observed. Because the deviation from spherical symmetry grows systematically with rotation frequency, it is essential to incorporate the actual rotation rate of observed neutron stars when deriving astrophysical constraints.

\section{Summary and Conclusions}

In this contribution, we have reported the results of KEH calculations for rapidly-rotating neutron stars with Gogny-type equations of state (EoSs). We have pointed out that the six Gogny EoSs except D1M$^*$ (\textit{i.e.}, D1, D1M, D1N, D1S, D1P, and D280) provide nonphysical results, predicting the pure neutron matter at certain high density region. Based on the KEH method, we have shown that central density (equatorial radius) is reduced (enlarged) by rotational effects, whose significance increases with rotational frequency, especially for $\Omega\gtrsim400$\,Hz.

In Ref.~\cite{Kwon2025a}, we demonstrated that rotational effects cannot be ignored for frequencies above 400\,Hz and calculations were performed at spin frequencies corresponding to observed pulsars: PSR J0030+0451 \cite{Miller2019}, PSR J0740+6620 \cite{Wolff2021}, and PSR J1748–2446ad \cite{Hessels2006}. For rotation rates below 400\,Hz, the difference from spherical models was small and did not significantly affect the constraining results. However, at 716\,Hz, we found that neutron stars could exist below the previously suggested constraint line. Those results are consistent with the results reported in this paper.

Looking ahead, in Ref.~\cite{Kwon2025b}, we discussed the relationship between the slope of the symmetry energy $(L)$ near saturation density, the neutron star radius, and the rotation-induced change in radius. We plan to pursue further systematic investigations to understand the relationship between neutron star structure and nuclear matter EoSs.

\begin{bibliography}{reference}

\end{bibliography}

\end{document}